\documentstyle[12pt,epsfig]{iopart}

\def \be {\begin{equation}}
\def \ee {\end{equation}}
\def \bea {\begin{eqnarray}}
\def \eea {\end{eqnarray}}

\def \vec #1{\mbox{\boldmath ${#1}$}}

\begin{document}

\title{Translational Invariance in Models for Low-Temperature 
Properties of Glasses}

\author{Reimer K\"uhn and Jens Urmann
}
\address{Institut f\"ur Theoretische Physik, Universit\"at Heidelberg\\
Philosophenweg 19, 69120 Heidelberg, Germany}

\begin{abstract}
We report on a refined version of our spin-glass type approach to the 
low-temperature physics of structural glasses. Its key idea is based on 
a Born von Karman  expansion of the interaction potential about a set of 
reference positions in which glassy aspects are modeled by taking the 
harmonic contribution within this expansion to be random. Within the 
present refined version the expansion at the harmonic level is reorganized
so as to respect the principle of global translational invariance. By 
implementing this principle, we have for the first time a mechanism that 
fixes the distribution of the parameters characterizing the local potential 
energy configurations responsible for glassy low-temperature anomalies solely 
in terms of assumptions about interactions at a microscopic level.

\end{abstract}



\section{Introduction}

The present contribution is intended to further explore and refine
our spin-glass way of looking at low-temperature anomalies in glasses
developed earlier \cite{Ku96,KuHo97}. The term `low-temperature anomalies'
(LTA) refers to a set of observations according to which a number of 
thermodynamic and transport properties of glassy and amorphous systems 
have been found to differ drastically and unexpectedly from those of their 
crystalline counterparts \cite{ZePo71}, when the temperature is lowered 
to a few degrees Kelvin. 

In particular, below approximately 1\,K the specific heat of glassy materials
has been found \cite{ZePo71} to scale approximately linearly with temperature, 
$C\sim T$, while the corresponding scaling for the thermal conductivity 
$\kappa$ is approximately quadratic, $\kappa \sim T^2$. Both findings contrast 
the $T^3$ behaviour of these quantities in crystals. Between 1\, K and 
approximately 20\,K the thermal conductivity displays a plateau and continues 
to rise as the temperature is further increased. The specific heat also 
changes its behaviour in the 1--20\,K regime. It exhibits a peak if displayed 
in $C/T^3$ plots, signifying an excess density of states in that energy range,
which is often referred to as the Bose peak. LTA are remarkable both for their 
ubiquity, and for their peculiar pattern of universality and absence thereof 
in various temperature ranges \cite{ZePo71,FrAn86}; for reviews, see 
\cite{HuAr,Phi87} and the recent collection \cite{Esq99}.

The anomalous properties of glasses at low temperatures are usually attributed 
to the existence of a broad range of localized low-energy excitations in 
amorphous systems -- excitations not available in crystalline materials.
At energies below 1\,K, these are thought to be tunneling excitations of 
single particles or small groups of particles in double--well configurations 
of the potential energy (DWPs). This is the main ingredient of the 
phenomenological so--called standard  tunneling model (STM), independently 
proposed by Phillips \cite{Phi72} and by Anderson et al. \cite{An+72}. As a 
second ingredient of the STM, it is supposed that the local DWPs in amorphous 
systems are random, and specific assumptions concerning the distribution of 
the parameters characterizing them are advanced to describe the experimental 
data below 1\,K \cite{Phi72,An+72}. Excitations at energies between 1\,K and 
20\,K responsible for the Boson peak and (via resonant scattering of phonons) 
presumably for the plateau in the thermal conductivity are believed to be 
of a different nature, namely localized vibrations in anharmonic single-well 
configurations of the potential energy (SWPs). The existence of such 
single-well potentials is the main additional assumption of the likewise 
phenomenological soft-potential model \cite{Ka+83,Bu+92} (SPM). Within the 
SPM it is supposed that locally the potential energy surface (along some 
reaction coordinate) can be described by certain fourth order polynomials, 
with coefficients distributed in a specific way so as to comprise both DWPs 
and SWPs, the former giving rise to tunneling systems, the latter to localized 
vibrations. 

In both, the STM and the SPM, a weak coupling between localized excitations
and extended (phonon) modes is assumed to describe the phenomenology of heat
transport and the anomalous acoustic or dielectric properties of glasses at 
low temperatures \cite{Jae72}.

Although the STM and the SPM describe the phenomenology of glassy LTA 
reasonably well, the situation cannot be considered entirely satisfactory.

To begin with, neither model accounts for a {\em mechanism\/} that would 
explain {\em how\/} the required local potential energy configurations 
would arise, and how they would do so with the required statistics. Indeed, 
as we have discussed elsewhere \cite{KuHo99}, the assumptions concerning 
the distributions of parameters characterizing the local potential 
energy configurations in either phenomenological approach are at best only 
partially plausible. Moreover, neither model can explain the considerable 
degree of universality of the LTA, or the observed absence of universality 
at intermediate temperatures. As phenomenological models for LTA, they also 
have little to say about relations between low-temperature phenomena and 
the physics at the glass transition. Finally, there is growing experimental 
evidence that things may go wrong with the assumptions of 
the STM (and the SPM) as the temperature is lowered into the mK regime. Let 
us mention (i) the unexpectedly rapid decay of coherent echos in glasses 
\cite{En+96}, (ii) the temperature dependence of acoustic attenuation and 
dispersion \cite {Nat+98}, (iii) unexpected behaviour in the long-time 
behaviour of hole-burning experiments \cite {Mai+96}, (iv])the unreasonably 
large value of 3\,mK for the minimal tunneling matrix element deduced from 
specific heat data \cite{StMe99} and dielectric measurements \cite{Rog+97}, 
and finally (v) the recently reported evidence for a macroscopic quantum 
states of tunneling systems in glasses below 5\,mK \cite{Stre+98} -- all 
pointing towards the necessity for a better understanding of interaction 
effects in glasses at (very) low temperatures.

It is with these observations in mind that our model based spin-glass way
of looking at low-temperature anomalies in glasses attempts to fill a gap,
and we believe it to carry considerable potential to clarify at least
some of the unresolved issues that have emerged lately.

Our approach is based on a Born von Karman expansion of the interaction 
potential of a glassy system about a set of reference positions, 
in which glassy aspects are modeled by taking the harmonic contribution 
within this expansion to be random. We derive the justification for such a 
procedure from the observation of universality: since the low-temperature 
anomalies observed in amorphous systems are apparently to a large extent 
insensitive to the the detailed form of the interaction, {\em  any\/}
interaction might be taken as a starting point, as long as it does give
rise to a glassy low-temperature phase. 

The approach leads to a class of models of spin-glass type which 
exhibit both, glassy low--temperature phases, and double- and 
single-well configurations in their potential energy. The distribution 
of parameters characterizing the local potential energy configurations 
can be {\em computed\/}, and differ from those assumed in the standard 
tunneling model and its variants. Still, the low-temperature anomalies
characteristic of amorphous systems are reproduced, and we are able to
distinguish properties which can be expected to be universal from those
which cannot. 

We have organized our material as follows. In section 2 we briefly describe
the original variant of our approach and describe the main results derived
from it. Section 3 introduces a refined version in which the expansion at 
the harmonic level is modified to respect the principle of global translational 
invariance, from which we derive new insights concerning the relation 
between the original particle-particle interaction and the distribution
of parameters characterizing  the local potential energy configurations.
Section 4 is devoted to main results, and and Section 5 to concluding remarks.

\section{The spin-glass way -- original setup}
\label{sec:section1}

We begin by briefly describing the main ingredients of our original model,
referring to \cite{Ku96,KuHo97,KuHo99,HoKu99} for details, numerical results, 
and illustrations.

In the original variant of our spin-glass approach to glassy low-temperature
anomalies, we suggested to consider the following Hamiltonian as a candidate 
for the description of a set of particles forming a glassy system, 
\be
{\cal H} = \sum_{i=1}^N \frac{p_i^2}{2 m} + U_{\rm int}(\{u_i\})\ ,
\ee
with an interaction energy given by
\be
U_{\rm int}(\{u_i\}) = -\frac{1}{2} \sum_{i\ne j} J_{ij} u_i u_j + \sum_i G(u_i)\ ,
\label{uint}
\ee
in which glassy aspects are modeled by taking the the $J_{ij}$ to be random.
On--site potentials of the form
\be
G(u) = \frac{a_2}{2} u^2 + \frac{a_4}{4\,{\rm !}} u^4
\label{gv}
\ee
are included to stabilize the system as a whole. Through the harmonic term in 
$G$, the parameter $a_2$ controls the number of modes that  are unstable at 
the harmonic level of description. The parameter $a_2$ may be fixed, or chosen 
according to some non-degenerate distribution as well (see below).

The description is in terms of {\em localized\/} degrees of freedom, i.e., 
the $u_i$ are interpreted as deviations of particle positions from a given set 
of reference positions as in a a Born von Karman expansion known from the 
dynamical theory of crystalline solids. Thus, the system is assumed to be 
already in a solid state, and  no attempt to provide a faithful description 
of the liquid phase is made.

The random interactions are chosen in such a way that the system can be 
analyzed within replica mean-field theory, e.g., as in the SK model \cite{SK}.
It implies that the potential energy surface of the system can be represented 
as a sum of effective independent single-site potential energies $U_{\rm eff}
(u_i)$ containing random parameters,
\be
U_{\rm int}(\{u_i\}) \longrightarrow \sum_i U_{\rm eff}(u_i)\ .
\label{sumui}
\ee
In the replica symmetric (RS) approximations these potentials are of the form 
\cite{Ku96,KuHo97}
\be
U_{\rm eff}(u)  = - h_{\rm eff}\, u -\frac{1}{2}J^2 C u^2 + G(u)\ ,
\label{urs}
\ee
with
\be
h_{\rm eff} = h_{\rm RS} = J_0 p + J\sqrt{q}\, z
\label{hrs}
\ee
and $C=\beta(q_d-q)$. Here $p$ denotes a macroscopic polarization, and $q_d$
and $q$ the diagonal and off-diagonal entries of the RS Edwards-Anderson
order parameter matrix. Apart from randomness that may be present in $G(u)$,  
the effective single-site potentials contain a single random parameter, viz. 
the Gaussian distributed effective fields $h_{\rm eff}$ having mean $J_0 p$ 
and variance $J^2 q$. The parameters of $p$, $q$ and $C$ characterizing the 
$U_{\rm eff}(u)$ ensemble are determined self--consistently through a set of 
saddle point equations.\footnote{Note that we have changed notation in 
comparison with our earlier papers, and that we are also using slightly 
different conventions -- omitting in particular the parameter $\gamma$ 
introduced in\cite{Ku96,KuHo97} in favour of the interaction scale $J$. 
It should be no problem for the reader to translate results whenever desired.} 

The model exhibits non-ergodic low-temperature phases, which may be glassy
or polarized, depending on the parameters $J_0$ and $J$; the glass-transition
temperature (or the transition temperature into the polarized phase) depends
on $a_2$ as well. Replica symmetry breaking (RSB) occurs at low temperatures 
and small $a_2$, and has been analyzed within a one-step replica-symmetry 
breaking (1RSB) approximation in \cite{KuHo97}. Its main effect is to modify
the $h_{\rm eff}$--distribution, and in particular to reveal correlations 
between the $h_{\rm eff}$ and whatever randomness one might have considered 
for the $G(u)$.

The relevance of these results for glassy LTA derives from the fact that
the $U_{\rm eff}(u)$ acquire a harmonic term  $-\frac{1}{2} J^2C u^2$ -- 
entirely of collective origin -- that renormalizes the local harmonic restoring 
force produced by $G(u)$. Hence for
\be
J^2 C  > a_2
\ee
the total harmonic contribution to $U_{\rm eff}(u)$ becomes convex downward
near the origin $u=0$, so that for sufficiently small $h_{\rm eff}$ the 
effective single-site potential $U_{\rm eff}(u)$ attains a DWP--form,
which is of collective origin.

The existence within the ensemble of effective single-site potentials of a 
spectrum of DWPs with a broad distribution of asymmetries is mainly 
responsible for the appearance of glassy LTA via low-energy tunneling 
excitations with a virtually constant density of states (DOS) at low energies, 
giving rise to the well-known linear temperature scaling of the specific 
heat at low temperatures. Higher order excitations in DWPs and quasi-harmonic 
excitations in SWPs give rise to a peak in the DOS, and consequently to a Bose 
peak. These results were presented and discussed in detail in
\cite{Ku96,KuHo97,KuHo99}. 

Let us close this section with a few remarks concerning the distribution 
of parameters in the effective single-site potentials.

If we choose the parameter $a_2$ in (\ref{uint}), (\ref{gv}) to be fixed and
non-random, the effective single-site potentials $U_{\rm eff}(u_i)$ contain
only a single random parameter, namely the effective field $h_{\rm eff}$, in
contrast to the STM and the SPM, both of which assume two randomly varying 
parameters. Indeed, whereas the low-temperature specific heat comes out
correctly without a randomly varying harmonic contribution to $G(u)$, dynamic
properties such as ultrasound attenuation do require a broad $a_2$-distribution
in order to be reproduced correctly \cite{HoKu99,HoKu99b}. While an assumption
of this kind is natural within our approach in view of the fact that the 
harmonic contribution to $G(u)$ might have been omitted in favour of diagonal 
entries $J_{ii}$ in the (random) interaction matrix, it still creates the 
(awkward) need for an independent hypothesis concerning their distribution. 
We consider this awkward to a higher degree, as it affects {\em local\/} 
quantities, whereas assumptions about the $J_{ij}$-distribution for $i\ne j$  
contribute to the physics only via global {\em collective\/} effects less 
sensitive to details of the underlying assumptions. 

\section{Translationally invariant interactions}
\label{sec:section2}

We now describe a modification of the above setup in which the need for
an independent hypothesis concerning the distribution of {\em local\/} 
variables is avoided, while at the same time having the additional benefit
of providing an element of physics that had been missing in our original 
formulation.

To wit, the interpretation we have given in support of the ansatz (\ref{uint})
is that the first harmonic contribution originates from a Born von Karman 
expansion of the interaction energy about an (unknown) set of reference 
positions. Within such an interpretation, however, a bona-fide interaction
matrix ought to respect the principle of global translational invariance ---
a feature that had been missing in our original approach. One way of enforcing
this principle is to require $\sum_j J_{ij} =0 $ for {\em all\/} $i$, entailing
that the diagonal entries $J_{ii}$ of the interaction matrix are {\em not 
independent\/} of the off-diagonal ones: $J_{ii} = -\sum_{j(\ne i)} J_{ij}$.
In other words, translational invariance fixes the diagonal entries of the 
$J$-Matrix solely in terms of true interaction contributions. It turns out
that this rather slight modification has pronounced effects on the structure
of the theory, which we have only just begun to explore.

Investigating the consequences of this idea quantitatively we choose a 
slightly different formulation, replacing (\ref{uint}) by
\be
U_{\rm int}(\{u_i\}) = \frac{1}{4} \sum_{i,j} J_{ij} (u_i - u_j)^2 + \sum_i 
G(u_i)\ .
\label{uintn}
\ee
It has the property that a global translation $u_i \to u_i + u$ (for all $i$)
leaves the first contribution invariant, irrespectively of the choice of the
$J_{ij}$. As in \cite{Ku96,KuHo97} we take the $J_{ij}$ to be Gaussians, and 
we adhere to $G(u)$ being non-random and of the form (\ref{gv}), serving 
stabilizing purposes. The on-site potentials $G(u)$ do, of course, break
translational invariance. However, we have also begun to look at variants in 
which the stabilizing contributions are themselves chosen in a translationally 
invariant form; results will be presented elsewhere.

Within a replica mean-field analysis, we now need additional order parameters
beyond the polarization and the matrix of Edwards-Anderson order parameters to
characterize the collective properties of the system, namely three replica 
correlations $q_{abc} =N^{-1} \sum_i {\langle u_i^a u_i^b u_i^c\rangle}$, and 
four replica correlations $q_{abcd} =N^{-1} \sum_i {\langle u_i^a u_i^b u_i^c 
u_i^d\rangle}$, the former, however only for $a=b$, the latter either for 
$a=b=c=d$ or for $a=b$ and $c=d$.

So far we have studied the system only in a RS approximation. Details will be
presented elsewhere. Here we only state the main results. In RS one assumes 
$p_a = p$, $q_{aa}=q_d$ and  $q_{ab}=q$ for $a\ne b$ as before, and in addition
$q_{aaa}=R_d$, $q_{aaaa}=Q_d$, and $q_{aab}=R$, $q_{aabb}=Q$ for $a\ne b$.
It turns out that the four replica quantities cancel in RS expressions. The 
remaining order parameters  are given as solutions of
\begin{eqnarray}
&&p=\langle\, \langle u\rangle\, \rangle_{y,z}\ , \  \
q_d  = \langle\, \langle u^2\rangle\, \rangle_{y,z}\ , \  \
q   =   \langle\, \langle u\rangle^2\, \rangle_{y,z}\ , \nonumber\\
&& \\
&&~~~~~~ R_d= \langle\, \langle u^3\rangle\, \rangle_{y,z}\ , \  \
 R = \langle\, \langle u^2\rangle \langle u\rangle\, \rangle_{y,z}\ . \nonumber 
\label{fpersn}
\end{eqnarray}
Here $\langle\dots\rangle_{y,z}$ denotes an average over two uncorrelated 
standard Gaussians $y$ and $z$, while $\langle\dots\rangle$ is a thermal 
average corresponding to the RS single--site potential
\be
U_{\rm eff}(u) = - h_{\rm eff}\, u +\frac{1}{2} k_{\rm eff}\, u^2 + G(u)\ ,
\label{ursn}
\ee
with 
\be
h_{\rm eff} = J_0 p - \frac{1}{2}J^2 C_R + J p\, y + J\sqrt{q - p^2}\, z
\label{hrsn}
\ee 
and
\be
k_{\rm eff} = J_0 - J^2 C + J\, y\ ,
\label{krsn}
\ee 
with $C=\beta(q_d - q)$ as before, and $C_R = \beta(R_d - R)$. Note that the
structure of the theory is now no longer SK-like as in our original version.
In particular, by enforcing translational invariance at the harmonic level of the 
Born von Karman expansion, we now have two random variables characterizing the 
local effective potentials instead of one (on top of whatever local randomness 
one might wish to consider in the stabilizing on-site potentials $G(u)$).

\section{Main results}
\label{sec:section4}

We now turn to results. The main properties of the system are: (i) It exhibits
a transition into a glassy phase with $q\ne 0$ at low temperatures (and 
small $a_2$). (ii) At sufficiently large $J_0$, the transition is into a 
phase with macroscopic polarization $p\ne 0$. (iii) As in the original setup,
the transition into the glassy phase is continuous. (iv) In phases without
macroscopic polarization we have $R_d = R \equiv 0$. (v) Though we have not
yet performed the stability analysis, we expect RSB to occur at low 
temperatures (and small $a_2$).

With respect to the appearance of low-temperature anomalies, the following 
features deserve mention: (i) The potential energy landscape as represented
by the ensemble of effective single-site potentials (\ref{ursn})-(\ref{krsn})
exhibits DWPs and SWPs. (ii) The ensemble is now characterized by {\em two\/} 
random variables, a random effective field $h_{\rm eff}$, and a randomly 
varying contribution $k_{\rm eff}$ to the harmonic force constant. As a 
consequence, DWPs will occur with a broad distribution of asymmetries and 
barrier heights, and both `soft' and `hard' SWPs will be observed. (iii) In 
phases with $p\ne 0$ the effective field $h_{\rm eff}$ and the harmonic force 
constants $k_{\rm eff}$ are correlated already in RS; we expect correlations 
between them to emerge irrespective of the value of the macroscopic 
polarization  through RSB effects, much like in the original setup 
\cite{HoKu99b}.

At this point it is perhaps appropriate to connect our results with those of 
simulations performed to locate and characterize DWPs in Lennard Jones systems 
\cite{HeuSi93}. DWPs observed in such systems can be parameterized (along 
the reaction path) by fourth order polynomials of the form $U_{\rm DWP}(u) = 
d_1 u + d_2 u^2 +d_4 u^4$ (or some equivalent form obtainable through shifts of
coordinates). It is to be noted that such simulations will not have access to 
the full range of $(d_1,d_2,d_4)$-triples owing to the non-linear constraint 
$d_2^3 +\frac{27}{8} d_1^2 d_4 \le 0$ which characterizes DWPs, making it 
difficult to estimate the full distribution or to decide with confidence 
whether empirical correlations observed between the parameters are solely due 
to the nonlinear constraint that defines the data set or of deeper origin. 
For the model considered in the present paper, we have the chance to compare 
such numerical results with analytic predictions, e.g. with the marginal 
distributions of $d_1$ and $d_2$ {\em conditioned on finding DWPs}. To compare 
with the theory, we have to identify $d_1=-h_{\rm eff}$, $d_2= \frac{1}{2}(a_2 
+ k_{\rm eff})$ and $d_4 = \frac{a_4}{4\,{\rm !}}$. If we assume, for instance,
a non-random $d_4$ and fix the energy scale by choosing $d_4=1$, and 
furthermore specialize to $J_0=0$, entailing $p=R_d=R=0$, we obtain the 
following conditioned marginal probability densities,
\be
\hskip -1cm p(d_1 | {\rm DWP}) = \frac{{\rm const}}{\sqrt{2\pi J^2 q}} \,
\exp\left(-\frac{1}{2}\frac{d_1^2}{J^2 q}\right)
\,{\rm erfc}\left(\frac{3|d_1|^{3/2}-J^2 C}
{J\sqrt 2} \right)
\ee
and
\be
\hskip -1cm p(d_2 | {\rm DWP}) = \frac{{\rm const}}{\sqrt{2\pi J^2/4}}\,
\exp\left(-\frac{1}{2}\frac{[d_2-\frac{1}{2}(a_2 -J^2 C)]^2}{J^2/4}\right)
\,{\rm erf}\left(\frac{\sqrt{-\frac{8}{27} d_2^3}}
{\sqrt{2 J^2 q}} \right)\ .
\ee
These are compared with simulation results in Figure 1. For $d_1$ the agreement
is quite good, though not for $d_2$. Remaining discrepancies are more likely 
due to RSB effects (which induce a-priori correlations not related to the 
non-linear constraints) than to finite size effects. Indeed, a major effect
of RSB is expected to be a reduction of the value of $C$ \cite{KuHo97}, which
could already account for much of the observed discrepancy in the 
$d_2$-distribution. It should be interesting to see, how far our results might 
eventually carry to help rationalizing the findings in Lennard Jones systems.

\begin{figure}[htb] 
{\centering 
\epsfig{file=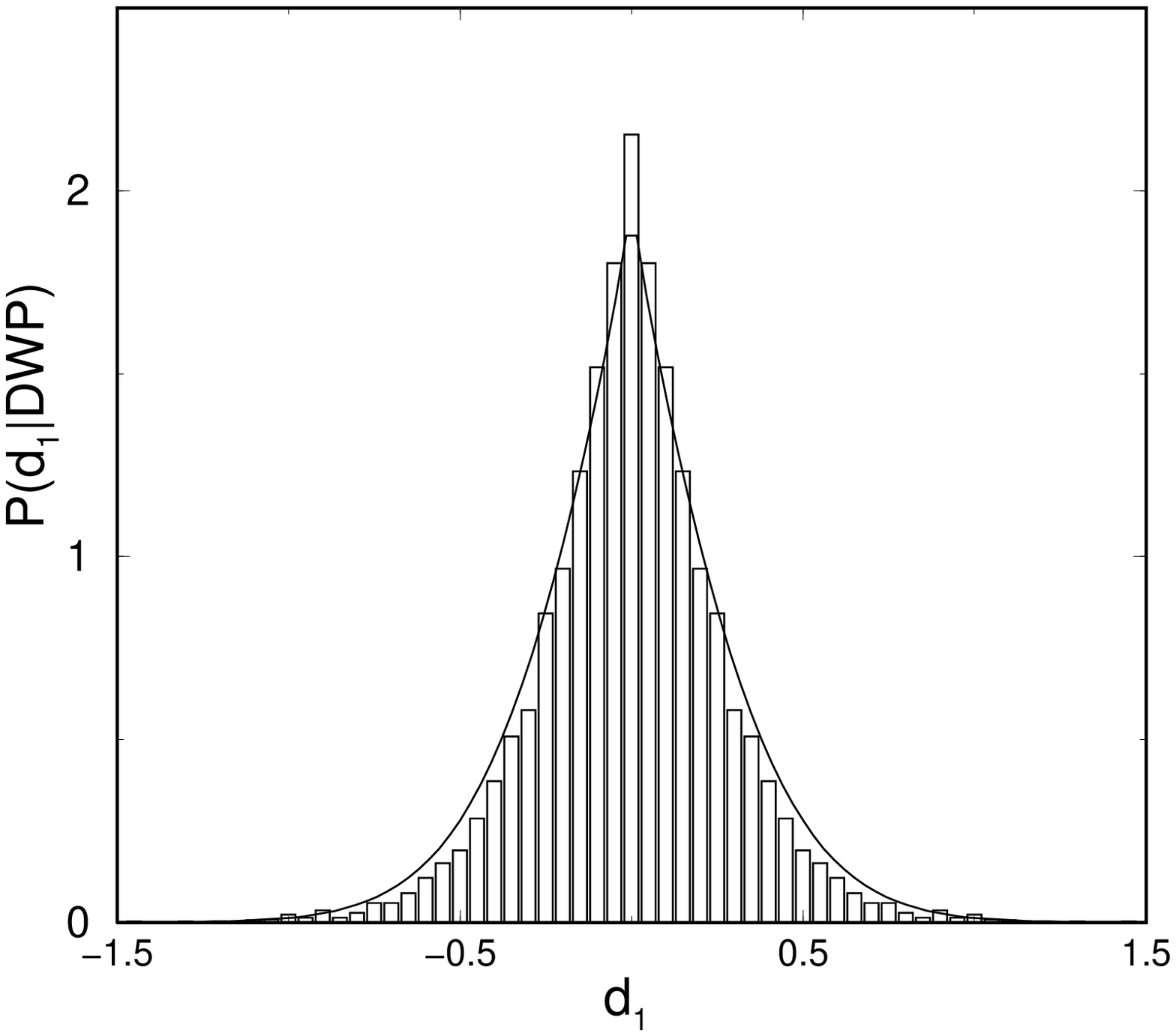, width=0.48\textwidth, height=5cm}  
\hfill{}
\epsfig{file=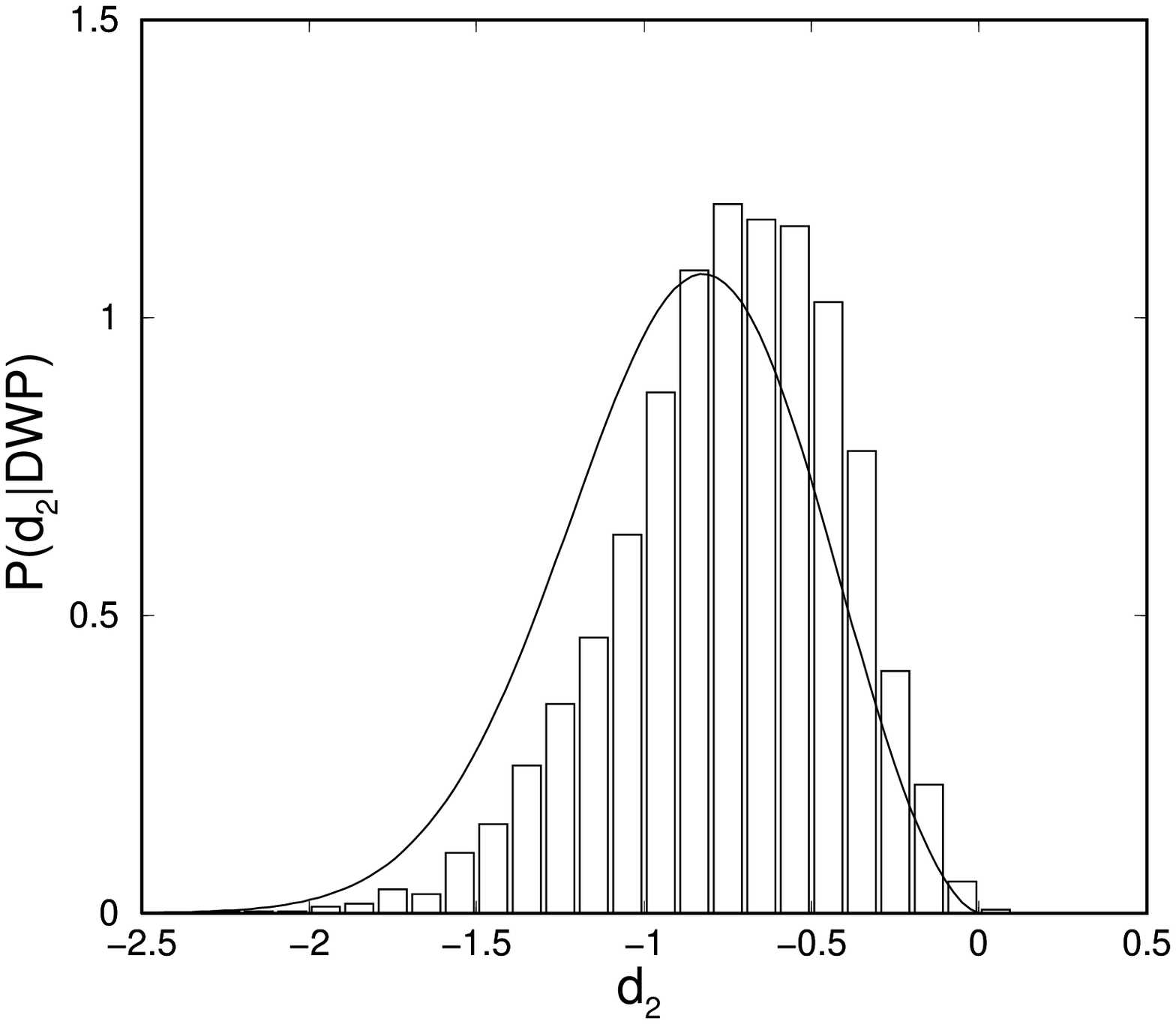, width=0.48\textwidth, height=5cm} 
\caption{Marginal probability densities for the parameters $d_1$ (left)
and $d_2$ (right) in an ensemble of DWPs generated by the interaction energy 
(\ref{ursn}), with $J_0=a_2=0$, $J=1$. and $d_1=a_4/4\,{\rm !}= 1$. Full 
lines: analytic prediction in RS. Simulation results were obtained from systems
of size $N=100$, using 1000 realizations so as to get reasonable statistics.} 
\par}
\label{f.fp} 
\end{figure}

Continuing the list of results with a bearing on LTA: (iv) As DWPs in the 
glassy phase (and in polarized phases), occur with a broad spectrum of 
asymmetries (generated by the $h_{\rm eff}$-distribution) they give rise to a 
broad spectrum of low-energy tunneling excitations with a nearly constant 
DOS, and thereby to the universal low-temperature anomalies of specific 
heat and thermal conductivity. (v) As in the original setup, higher order 
excitations in DWPs and quasi-harmonic excitations in SWPs occur with a 
peaked DOS, producing a Bose peak. (vi) Because of the unbounded $k_{\rm 
eff}$-distribution (Gaussian in RS), we do no longer see a possibility of 
having amorphous phases {\em without\/} DWPs, unlike in the original setup. 
(vii) Our previous analysis \cite{KuHo97,KuHo99}, as to which low-temperature 
properties might be universal (those related to tunneling excitations) and 
which not (those related to Bose peak phenomena), remains unaffected by the 
modifications of the present setup. (viii) Last but not least, the proposal 
(\ref{uintn}) embodies for the first time a {\em mechanism to generate a 
glassy potential energy landscape with an ensemble of SWPs and DWPs entirely 
in terms of interactions at the microscopic level via collective effects}, in 
which the distribution of both asymmetries and barrier heights of DWPs is 
amenable to analytic characterization and follows solely from assumptions 
about microscopic interactions. We regard this general qualitative result as 
perhaps one of the most significant advances contained in the present 
contribution. Indeed, whereas there have always been plausible arguments 
concerning relevant features of the distribution of asymmetries of DWPs 
responsible for the LTA (broad, symmetric smooth, hence nearly constant in the 
interesting energy range), such arguments have been missing for the 
distribution of barrier heights (or tunneling matrix elements); within 
phenomenological modeling, it was always necessary to {\em guess\/} these 
distributions in a way that gets the main physics correct. Within our refined 
approach there is no longer any need for guesswork at this level; everything 
follows from assumptions about the interactions at the microscopic level via 
collective effects (and should -- in view of the universality of glassy LTA -- 
not critically depend on details of these assumptions). We are, of course, 
aware that this statement ought to be checked in greater detail than we have 
so far been able to do.

\section{Concluding remarks}
\label{sec:section5}

Whether the recent problems with the interpretation of some experiments at 
very low temperatures which we mentioned in our introduction are related to 
the fact that the above-mentioned guesses concerning tunneling matrix elements 
(or concerning the absence of correlations between parameters characterizing 
local potential energy configurations) have not been entirely correct, or 
whether these problems point to more fundamental issues, we can at present 
not tell, as we have not yet been able to address these problems within the 
refined formulation of our spin-glass approach presented above.

The fact that the possibility for having phases without DWPs virtually
disappears when global translational invariance is taken into account within
a Born von Karman expansion of the interaction energy, casts some doubt on
the interpretation of the low internal friction results of Liu et al. 
\cite{Liu+97} as being indicative of an amorphous system  {\em without\/} 
low-energy tunneling excitations. While the density of such excitations may 
indeed be unusually low in the a-Si samples of Liu et al., amorphous systems 
without DWPs appear to be extremely unlikely within our new perspective. This 
issue certainly deserves deeper investigation.

Among the interesting problems within reach of our approach we might mention
in particular (i) an analysis of collective quantum effects \cite{Stre+98} via 
a full-fledged quantum statistical treatment of our spin-glass type models 
\cite{KuSh99}, (ii) a deeper understanding of relations between LTA and the 
phenomenology at the glass transition, (iii) an investigation of potential 
energy landscapes which respects the three-dimensional nature of deviations 
from reference positions in view of possible consequences for magnetic field 
effects and Aharonov-Bohm phases in glasses \cite{Ket+99}, replacing (\ref{uintn})
by $U_{\rm int}(\{u_i^\mu\}) = \frac{1}{4} \sum_{i,j} \sum_{\mu,\nu} J_{ij}^{\mu\nu}
(u_i^\mu - u_j^\mu) (u_i^\nu - u_j^\nu) + \sum_i G(\vec u_i)$, in which $\mu$ and 
$\nu$ label the three Cartesian components of the $\vec u_i$, (iv) the
investigation of dynamic effects in the vicinity of the glass-transition. In this
respect finally note that -- the glass transition in the present setup being 
continuous -- there is still need for improvements on the modeling side.

\medskip\noindent
{\bf Acknowledgments} We are indebted to A. Heuer, U. Horstmann, and  W. 
Schirmacher for very useful discussions.

\end{document}